\documentclass[letterpaper, 10pt, conference]{ieeeconf} 

\usepackage{booktabs,amsfonts,algorithmic,algorithm2e,amsmath,mathtools,framed,pbox,graphicx,amssymb,color,mathrsfs,amssymb}

\IEEEoverridecommandlockouts
\usepackage{cite}
\usepackage{graphicx}
\usepackage[bbgreekl]{mathbbol}

\DeclareMathSymbol\bbDelta  \mathord{bbold}{"01}

\usepackage[yyyymmdd]{datetime}

\long\def\/*#1*/{}

\title{\LARGE \bf Internal Feedback in Biological Control: \\ Architectures and Examples 

\thanks{$^1$Computation and Neural Systems, $^2$Computing and Mathematical Sciences, Caltech; $^3$Engineering Physics, KTH Royal Institute of Technology; $^4$HHMI Janelia Research Campus; $^5$Molecular Genetics and Cell Biology, $^6$Neurobiology, University of Chicago; $^7$Computational Neurobiology Laboratory, The Salk Institute for Biological Studies; $^8$Biological Sciences, University of California San Diego;. To whom correspondence may be addressed: {aasarma,
doyle}@caltech.edu}
\thanks{This paper is one of three in a series on internal feedback in biological control architectures. These papers may be read in any order, though a suggested order is this paper, then \cite{Paper2}, then \cite{Paper3}}

}
\author{Anish A. Sarma,$^1$ Jing Shuang (Lisa) Li,$^2$ Josefin Stenberg,$^3$
Gwyneth Card,$^4$\\ Elizabeth S. Heckscher,$^5$ Narayanan Kasthuri,$^6$ Terrence Sejnowski.$^{7,8}$ 
and John C. Doyle$^2$} 
\begin{document}
\maketitle
\noindent 

\textbf{Feedback is ubiquitous in both biological and engineered control systems. In biology, in addition to typical feedback between plant and controller, we observe feedback pathways \textit{within} control systems, which we call \textit{internal feedback pathways} (IFPs), that are often very complex. IFPs are most familiar in neural systems, our primary motivation, but they appear everywhere from bacterial signal transduction to the human immune system.
In this paper, we describe these very different motivating examples and introduce the concepts necessary to explain their complex IFPs, particularly the severe speed-accuracy tradeoffs that constrain the hardware in biology. We also sketch some minimal theory for extremely simplified toy models that nevertheless highlight the importance of diversity-enabled sweet spots (DESS) in mitigating the impact of hardware tradeoffs. For more realistic models, standard modern and robust control theory can give some insights into previously cryptic IFPs, and the new System Level Synthesis theory expands this substantially. These additional theories explaining IFPs will be explored in more detail in several companion papers.}

\section{Introduction}
Although it is common to observe similarities between engineered and biological complexity, it has remained challenging to import engineering theory into biology, or to import biological principles into engineering design, beyond special cases \cite{Lestas2010,Chandra2011,Wolpert1995,Aoki2019}. This motivates a theoretical effort to re-visit the assumptions in engineered control.

In biology, control (sensing, communications, computing, and actuation) is necessarily implemented in  components and networks that are sparse, local, delayed, quantized, noisy, nonnegative, saturating, etc, or \textit{SLD+} for short. In most control engineering, fast and cheap digital electronics make SLD+ constraints negligible in the communications and computing components, which is fortunate as such constraints made methods from modern control theory scale extremely poorly \cite{Witsenhausen1968, Anderson2019}. SLD+ constraints will also be an increasing challenge in engineering cyber-physical systems. These limits are especially severe in studying large biological or cyber-physical networks, so it is timely to revisit them.
\begin{figure*}
    \centering
    \includegraphics[width=\textwidth]{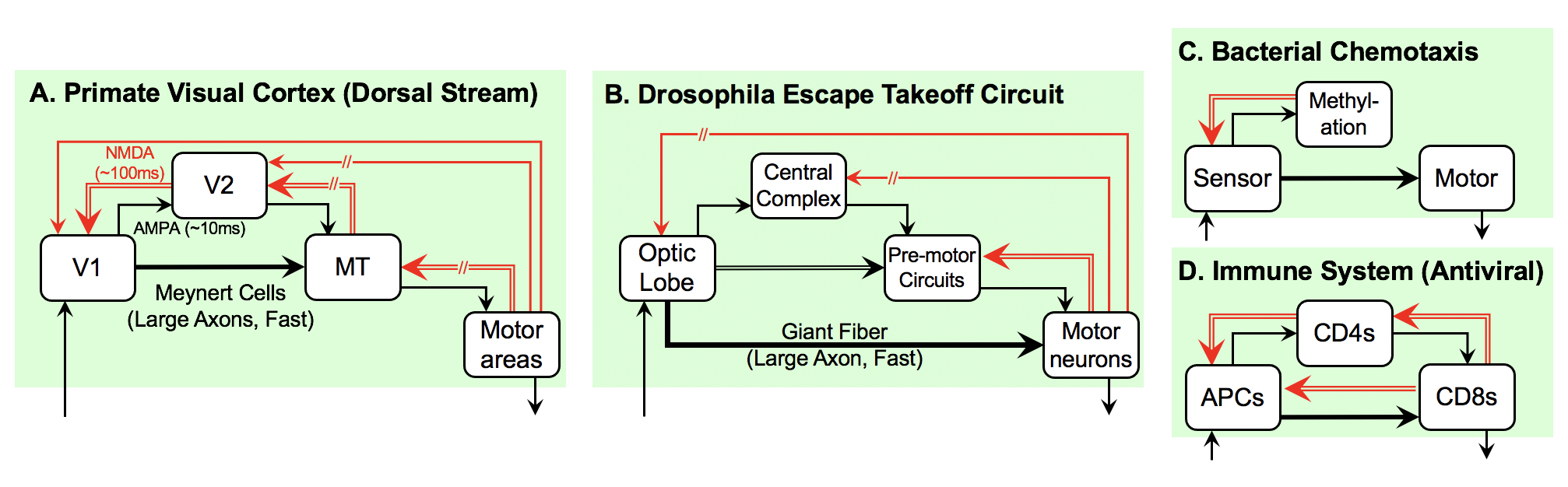}
	\caption{\textbf{Schematic layered architectures in biology with forward paths in black and IFPs in red.} \textbf{(A)} A simplified diagram of the motion processing circuit in primates, whose forward path has evolved to achieve time-sensitive tasks like predator evasion and prey tracking. Note the existence of multiple complex IFPs, which are believed to be essential for visual processing. Broken lines indicate additional synapses, not shown. It has been difficult to make further progress in understanding exactly why IFPs are helpful for these tasks because of the lack of a theoretical framework. \textbf{(B)} An escape circuits in adult flies, again with prominent fast forward paths and complex IFPs in a similar configuration. \textbf{(C-D)} Additional circuits from bacterial chemotaxis and a simplified immune antiviral response with similar IFP architectures.\label{fig:intro_figure_neuro}} 
\end{figure*}

We conjecture that the cryptic complexity in biological control, as seen in Fig. \ref{fig:intro_figure_neuro}, is largely due to SLD+ component constraints rarely present in engineering. 
A first step towards understanding this complexity \cite{ND15,Nakahira2021, Nakahira2019b} uses simple models and theory of a mountain biking video game experiment to introduce some essential concepts. 
The game has tunable requirements on player performance which, together with speed-accuracy tradeoffs (SATs) in sensor and actuator level components, require layered architectures in the nervous system to create diversity-enabled sweet spots (DESS) \cite{Nakahira2021, Nakahira2019b}. In this and companion papers \cite{Paper2,Paper3}, we extend these concepts to the more challenging SLD+ constraints on communication, computing, and control architecture, and provide new detailed explanations for why the \textit{internal feedback pathways} (IFPs) in red in Fig. \ref{fig:intro_figure_neuro} are not only plausible but necessary.

The new theory of System Level Synthesis (SLS) \cite{Anderson2019} greatly expands the scalability of control theory for SLD+ constraints, which we use to show \cite{Paper3} that the greatest source of IFPs is SLD+ in communications. In retrospect, some problems with SLD+ constraints in only sensing and actuation can also be solved with standard preSLS control theory which also leads to IFPs, though  much less complex than for SLD+ communications\cite{Paper2, Paper3}.

We briefly consider motivating case studies from three domains of biological complexity: bacterial cell signaling, neuroscience, and immunity. These domains, especially the latter two, are known for their sprawl of component types and system-level flexibility, which creates a need for theory but also a high barrier for theorists to generate and engage with suitable abstracted systems. To facilitate connections with control theory across varied domains in biology, we show that particular types of problems recur across these domains. These problems are familiar in their general shape to control theorists, but sufficiently distinct that they have not been well-studied in sufficient detail. We present simple descriptions here in the hopes of facilitating more collaborations as well as motivating results in companion papers\cite{Paper2,Paper3}.

The core concept and mystery to be explained in this paper and its companion papers \cite{Paper2,Paper3} is the \textit{internal feedback pathway} or IFP. Biology, like engineered control, is full sense-compute-actuate feedback loops. However, unlike engineered systems, at least as represented in standard theories, biology is also rich with examples of communication signals going in the opposite direction of a conventional feedback loop, such as communication from actuators to sensors. While the simplest cases of this type of IFP result from biological instantiations of typical control functions like estimation (and have been understood as such in the literature \cite{Wolpert1995,LeeMumford2003,Swain07}), many cases of IFPs in biological control settings appear to be unlike their counterparts in engineering. Fig. \ref{fig:intro_figure_neuro} shows some examples of the enormous amount of IFPs present in biological systems, unexplained by traditional theory. It is also a highly simplified cartoon that hides enormous additional complexity in lower level cells and signaling molecules.

\section{Concepts and Architectures}\label{section2}

\subsection{Supporting concepts}
Four concepts are essential to our study of IFPs. We discuss these in the context of the simple problem from neuroscience introduced in \cite{ND15} and developed further in \cite{Nakahira2021}, a problem which by design highlights the concepts necessary to understand DESS. Although IFPs are explicitly not a part of this earlier theory, they follow naturally from its main concepts.
\begin{figure*}
\begin{minipage}{.6\textwidth}
\begin{flushleft}
\includegraphics[width=0.95\textwidth]{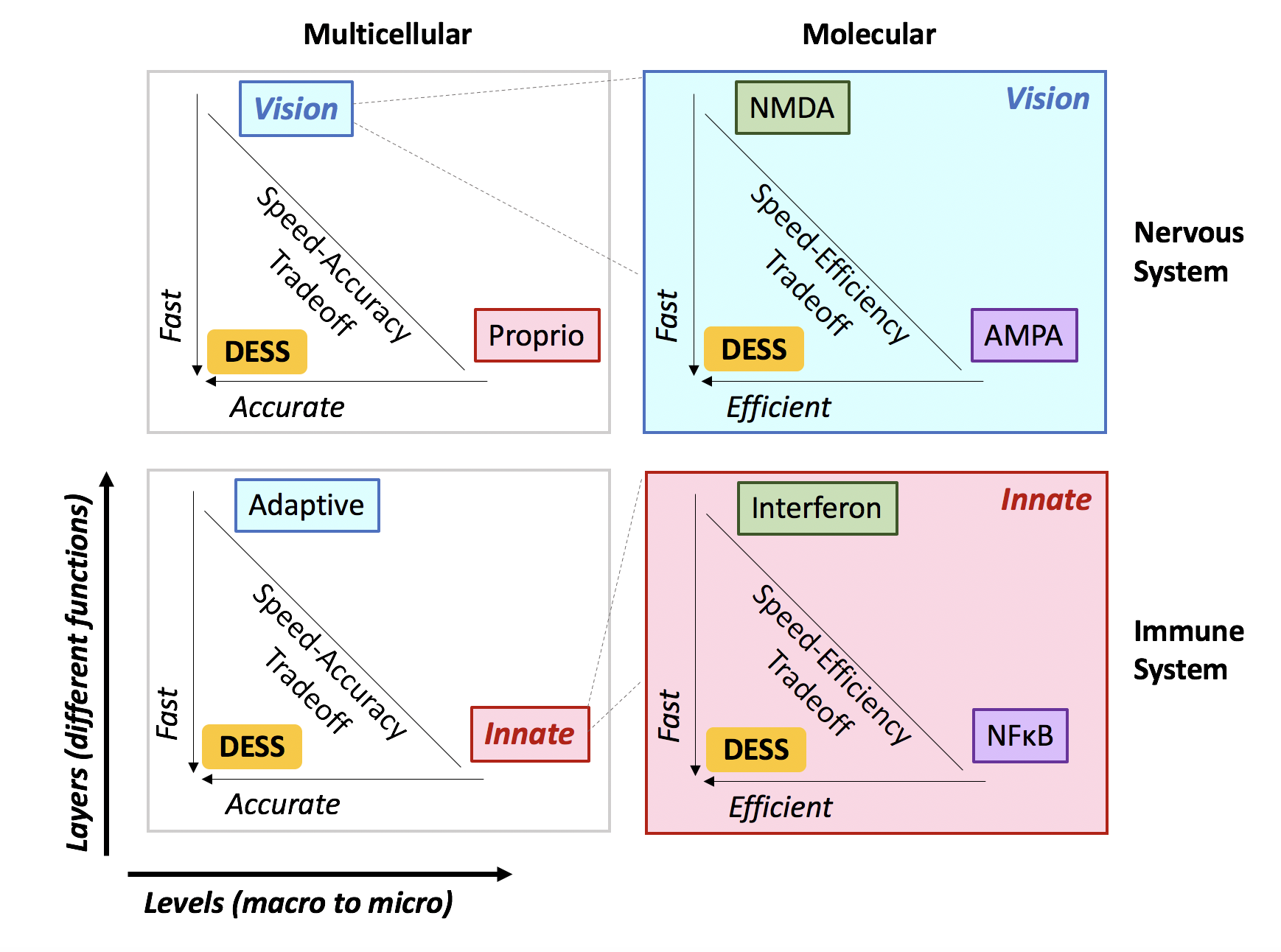}
\end{flushleft}
\end{minipage}
\begin{minipage}{.4\textwidth}
\begin{flushright}
\caption{\textbf{Layers, levels, tradeoffs, and diversity-enabled sweet spots.} Biological system complexity can be interpreted through the framework of layered control architectures. We consider two systems: the nervous system (top row) and the immune system (bottom row). At a given \textit{level} of analysis (multicellular on the left, molecular on on the right), individual components face tradeoffs between crucial system-level goals like speed, accuracy, efficiency, and robustness. These tradeoffs are interrelated, but we consider two dimensions at a time for the sake of schematic portrayals. Both the nervous system and the immune system are layered, meaning they exhibit a hierarchy of specialized control loops. In both cases, the components in the layered architecture exhibit a speed-accuracy tradeoff. A faster and less accurate layer can be combined with a slower and more accurate layer to achieve fast and accurate system-level performance, a combination we call a diversity-enabled sweet spot DESS) \cite{Nakahira2021}. Within each system or cell-level component is a molecular level, which itself has diverse components and tradeoffs that are reflected at the system level; here we focus on speed-efficiency tradeoffs, with faster signaling in both cases requiring specialized high-turnover molecular signaling systems. \label{fig:intro_figure_layers}}
\end{flushright}
\end{minipage}

\end{figure*}
The core problem from \cite{ND15,Nakahira2021} is motivated by multisensory tasks like mountain biking, which involve diverse sensing, control, and actuation components. In the simplified problem, the mountain-biking task naturally separates in theory and experimental data into a ``trail'' layer of reference tracking with advance warning (trail following using vision) and a ``bumps'' layer that does disturbance rejection (quickly responding to bumps in the terrain, using proprioceptive and other reflexes). 

\textit{Levels.} The most familiar concept is that biology has distinct levels, or scales (e.g. from molecules to synapses, cells, and circuits). In \cite{Nakahira2021} there were two levels: systems (trail and bumps layers) and their implementation in nerves. In electronics, a typical example of levels would be between a model of semiconductor physics in transistors and a circuit diagram that includes amplifiers. Fig. \ref{fig:intro_figure_layers} shows systems and molecule levels, skipping nerves and cells.

\textit{Speed-accuracy tradeoffs (SATs).} Biological hardware has severe speed-accuracy tradeoffs (SATs), another familiar concept that is expanded and extended in our theory. In computers, different types of storage (registers, cache, RAM, disk, etc.) have different speeds of access and storage capacities. In neuroscience, vision is slower and more accurate than proprioception. In immunology, adaptive immune responses take several days longer to mount than innate immune responses, but adaptive responses are more targeted.

\textit{Diversity-enabled sweet spots (DESS).} 
If biological systems were build out of homogeneous components, the extreme SATs in nerves and the immune system make robust control seemingly impossible. However, these SATs allow for extreme diversity in the hardware, which can be leveraged with the right architectures.
Highly diverse hardware-level components (which are constrained by SATs) enable performance sweet-spots which largely overcome the severe hardware-level SATs. In computers, such sweet spots include virtual memory management systems. In neuroscience, extreme diversity in axon sizes, receptors, and neurotransmitters are abundant \cite{Nakahira2021}, but largely hidden in Fig. \ref{fig:intro_figure_neuro}. By itself, diversity of components only enables sweet spots; to achieve these sweet spots requires specific architectures to maximize the utility of each component, which we call DESS. 

\textit{Layered architecture.} Layered architectures (e.g. planning and reflex, vision and proprioception, adaptive and innate, integration and change-detection, software and hardware) help create DESS \cite{Nakahira2021, Nakahira2019b}. We claim that DESS is a universal purpose of complex architectures, but architecture itself remains a largely ad hoc and domain-specific subject. Fig. \ref{fig:intro_figure_layers} highlights the essential concepts of layered architecture, SATs, and DESS. For the nervous system case there is also simple theory and associated experiments \cite{Liu2019, Nakahira2021}. Naively, success in the biking task seems to require speed and accuracy that the raw hardware lacks, making non-layered solutions infeasible. The layered nervous system breaks the overall biking problem into a high \textit{trails} layer of slow but accurate vision with trail look-ahead for advanced warning, and a low \textit{bumps} layer that uses fast but inaccurate muscle spindles and proprioception to sense and reject bump disturbances. The motor commands from these two control loops to the muscles simply add in the optimal case, as well as in experiments \cite{Nakahira2021,Liu2019}. Effective architectures (such as layering) create a DESS where diverse hardware enables a sweet spot that is both fast and accurate. 

\subsection{Architecture: SATs, DeSS, Layering, and Theory}

In this section, we review the theory in \cite{Nakahira2021} to show how levels, SATs, DeSS, and layering can be captured in a minimal model, initially without internal controller constraints or IFP. Then in a companion paper \cite{Paper2} we'll focus on a single layer theory with SATs on internal sensing and actuation and where DESS necessitates IFP. SATs on communications requires SLS theory and results in much more complex IFP\cite{Paper3}.

\begin{figure}
    \centering
    \includegraphics[width=\columnwidth]{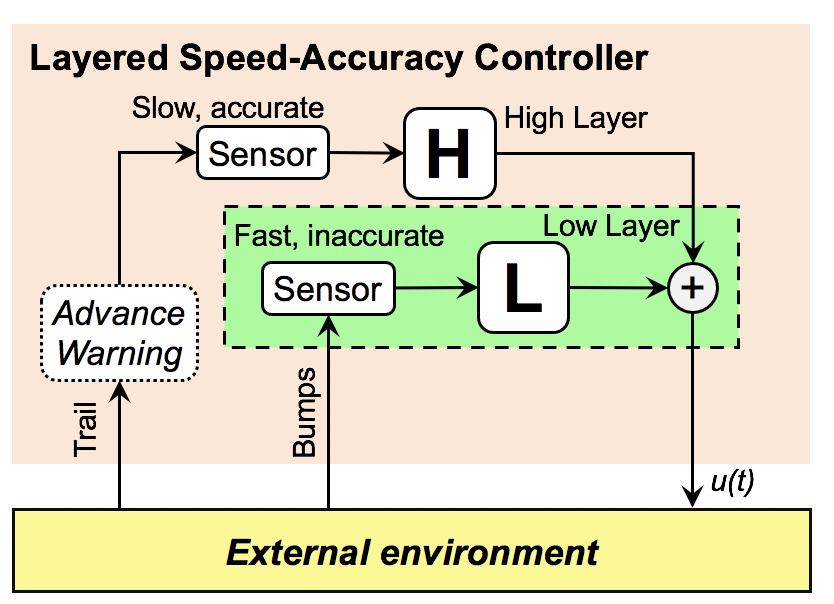}
	\caption{\textbf{A layered architecture making use of from diverse components.} Layered controller adapted from a model of bike riding\cite{Nakahira2021} that illustrate DeSS. This controller optimally layers diverse speed- and accuracy-limited components from axons to muscle to produce optimal behavior with good speed \textit{and} accuracy in both theory and experiments. The lower layer (L) uses a more quantized sensor with small delays, wile the higher layer controller (H) uses a delayed sensor with less quantization. The optimal use of the two control loops together is for their commands to simply add at the actuator.\label{fig:ynfig}} 
	
\end{figure}

A minimal version of \cite{Nakahira2021} with a scalar neutrally stable plant and two control layers or loops is
\begin{equation}
    \begin{gathered}
        x(t+1) = x(t)+u(t)+w(t)
        \\w(t)=v(t)+r(t-T_r)
        \\u(t) = u_{L}(t)+u_{H}(t)
        \\u_{L}(t)= L(Q_L(x(1:t-T_L)))
        \\u_{H}(t)= H(Q_H(r(1:t-T_H)))
    \end{gathered}
    \label{eq:YorieM}
\end{equation}
where $Q_L$ and $Q_H$ are quantizers, $T_L$ and $T_H$ are delays, and $L$ and $H$ are controllers. The $L$ and $H$ refer to low and high layers, respectively. This is shown schematically in the diagram in Fig. \ref{fig:ynfig}.

In \cite{Nakahira2021}, this models a mountain biking game where $r(t)$ is the trail and $v(t)$ models bump disturbances. The trail is tracked via $u_H$ and $K_H$ using direct visual sensing of $r(t)$ of the form $Q_H(r(1:t-T_H))$. The $T_r$ delay in $r(t-T_r)$ models a look-ahead advanced warning of $T_r$, so $T_r-T_H$ is the net advanced warning. The lower layer reflex controller for bumps is $u_L$ and $L$ which senses the state $x(t)$ with delay $T_L$ and quantization $Q_L$. In standard control terminology, $L$ would be feedback and $H$ would be called feedforward, but we will avoid the term here.

Quantization and delay can be modeled with a variety of abstractions, which depends on neuron and nerve level models as well as coding schemes, but the results seem largely robust to assumptions\cite{Nakahira2021}, and are both dramatic and qualitatively intuitive. What is essential is a SAT so that similar size nerves range from many small axons (slow but accurate) to few large axons (fast but inaccurate). Given a variety of bio-plausible SATs, having extreme diversity between such nerves (in $H$ and $L$ layers) is far better than uniform nerves for eq. (\ref{eq:YorieM}). This is qualitatively consistent with known physiology (vision is slow but accurate, and reflexes are fast and less accurate) and now extensive experimentation \cite{Nakahira2021}, which further solves for optimal $Q$ and $T$ with a variety of assumptions. In all cases, the layering in eq. (\ref{eq:YorieM}) is optimal over all controller architectures, and each layer can even be solved separately. 

These results\cite{Nakahira2021} formalize all four of our essential concepts: \textit{levels} (system and nerve) to derive \textit{SATs} and dramatic \textit{DESS} due to \textit{layering}, a surprisingly rich outcome for such a simple experiment, model, and theory. However, important features of the system, such as IFPs within the visual system, are abstracted away as simple external delays, while IFPs to the the sensors and actuators are excluded by assumption. While some delays and quantization are plausibly modeled as external, many are not. In \cite{Nakahira2021}, both were external and optimized along with $H$ and $L$. This fit the video game experiment and greatly simplified both data and theory, but is not plausible in more realistic settings where the SATs in sense and act are internal and in some cases tunable, and optimal controllers have complex IFPs. The model in eq. (\ref{eq:YorieM}) has a scalar state, so realistic sparsity (another source of IFPs) cannot arise.

The simplest next step is to include vector and unstable dynamics with delays and sparsity on internal sensors and actuators, which results in extensive IFPs and can be solved with modest extensions to standard control theory\cite{Paper2}. The most general case allows delays, sparsity, and locality in internal communications, requires SLS, and has much more complex IFPs\cite{Paper3}. Next we use familiar biological case studies to motivate these more biologically plausible assumptions that then lead in theory to complex IFPs strikingly consistent with the motivating biology.

\section{Case Studies }
We now consider three case studies in greater detail. These case studies are somewhat dense, as they concern complex biological systems. However, our hope is that the preceding discussion will help readers new to the biology understand the key features of each system from a theoretical perspective. We note that each system is built out of quite different parts and performs different functions, so the fact that the simple architectural features above offer any insight suggests that some architectures recur in biology despite the superficial variation.

\subsection{Bacterial Chemotaxis}
Bacterial chemotaxis, a behavior in which bacteria swim up nutrient gradients, is a canonical case study in robust biomolecular signal transduction and control (typically studied in \textit{E. coli} \cite{Barkai97,Alon99}). Previous work has shown that the gradient-following behavior in bacteria is equivalent to step adaptation, which necessarily requires integral feedback \cite{Yi2000}. However, integral feedback might have several realizations, and beyond this, several biophysical \textit{implementations}. The implementation used by the chemotaxis system has rich structure that is not easily captured by state-space realizations.

Sensors for nutrients cluster at one end of the oblong bacterial body, called the \textit{nose}. At baseline, the nose is made up of receptors in a methylated state and the bacteria \textit{tumbles} randomly in place. When these receptors bind to a nutrient molecule, they set off two concurrent pathways within the bacterial cell. The forward motor pathway, via an ATP-intensive phosphorylation cascade, sets off flagellar rotation, sending the bacteria on a \textit{run} in the direction of the nutrient. The adaptation pathway, also ATP mediated but less costly, demethylates the receptors, making them less sensitive and the bacteria resumes tumbling. Higher nutrient concentrations are then needed to resume a run, ensuring that the bacteria will tend to swim in the direction of the gradient, and under light assumptions, will converge to the source of a diffusing nutrient across huge ranges of absolute concentrations. 

The main source of diversity and SATs in the system is apparent in the two signaling pathways. The timescales of low-energy methylation and high-energy phosphorylation differ by an order of magnitude \cite{Barkai97,Alon99}. Phosphorylation is fast, but single-bit: the individual bacterium either runs or tumbles. To more deeply understand this signaling network we can consider counterfactual implementations of the same function. For instance, the sign of the methylation and demethylation processes might be arbitrary in a theoretical model of integral feedback, but in the biological case, it is highly relevant: demethylation, which costs ATP, occurs when the bacterium is in nutrient-rich environments where ATP is easily replenished. Methylation, which occurs without ATP, allows the bacterium to relax to a more sensitized state. 

Another relevant biophysical detail is that the sensors are clustered on the nose of the bacterium, allowing accurate signals to be passed efficiently between them \cite{LevinBray02}. By contrast, the flagellar motors are necessarily distributed around the cell body to allow rotational movement, and signaling to these molecules must be both fast and global. A plausible alternate implementation would place the adaptation circuit at the motor, rather than the sensor; however, if this were the case, the system would be required to send ATP-costly broadcast signals with every nutrient binding event, rather than only spending ATP on changes in detected concentrations. Thus it is not just the existence of the IFPs but \textit{where} they are in the control loop and in the organism that enables the flexible, fast, and accurate behavior.

\subsection{Nervous Systems}

Previous work \cite{Nakahira2021} proposed a tradeoff principle for the organization of bundles of axons among the cranial nerves (which have roughly similar diameters) : in a given physical diameter, a cranial nerve could have many small axons or a few large axons. Large-diameter axons are faster, but packing in more axons has higher bandwidth, so cranial and other nerves face a speed-accuracy tradeoff.

Such diversity in neuronal morphology is widely observed in evolutionary contexts and directly related to speed at the single neuron level. However, these extreme cases have typically not been emphasized in theories (where neurons are often modeled as homogeneous); they have also not been connected to IFPs nor to the computation power and flexibility of cortex. Here, we argue that not only are extreme neurons and neuronal diversity crucial for understanding cortex, they are crucial for understanding IFPs as well. In cortex, the largest and most striking neurons are the large pyramidal cells, also called Meynert Cells in visual cortex and Betz Cells in motor cortex, which are involved specifically with conducting rapid moving-object changes in visual scenes or rapid responses to perturbations in planned movements, respectively. As with the Giant Fiber system in flies (Fig. \ref{fig:intro_figure_neuro}), these cells will necessarily require IFPs to achieve flexibility and accuracy while also retaining speed.

The Giant Fiber is important in escape behaviors in the fruit fly \textit{Drosophila melanogaster}. It is a unique neuron type specialized for rapid conduction: the large-axon Giant Fiber. The Giant Fiber integrates processed visual signals from the fly's retina-like optic lobe and projects to the nerve cord, where it synapses directly on the motor neuron that drives the ``jump" muscle. The timing of a single spike in the Giant Fiber determines whether the fly will choose speed (fast takeoff) or accuracy (slow takeoff) in its escape \cite{vonReyn2017, Ache2019_looming, vonReyn2014}. The Giant Fiber is thus part of a very direct path from photoreceptor to muscle, for which we know the identity of each neuron and, at least roughly, the computation performed across each synapse. This pathway participates in a multi-layered visual processing system in which component diversity (axon diameter) is a prominent feature. The theory described above and in \cite{Paper2,Paper3} offers an initial theoretical explanation not only for what the Giant Fiber system should do but also for what information must be carried by IFPs to the dendrites of the the Giant Fiber in order to make rapid responses possible.

\subsection{Immunity}

The immune system can roughly be divided into two layers: innate and adaptive immunity \cite{Janeway}. Components are quite diverse both within each layer and between layers. Innate cells exhibit more morphological specialization (e.g. macrophages for absorbing debris or neutrophils for releasing bactericidial granules). Innate immunity provides fast, somewhat inaccurate protection, in minutes to hours, but is less specific to pathogens than adaptive immune immunity. 

Adaptive immune cells are less morphologically distinct but much more specialized at the molecular level, with tens of thousands of unique receptors in the repertoire of T cells and B cells. These receptors take time to generate and select (through fascinating mechanisms that we elide here for space), making adaptive immunity accurate but slow. Targeted protection against pathogens is achieved only after several days. Once available, adaptive immunity clears pathogens much more effectively than innate cells alone -- often by coordinating more targeted responses on the part of innate immune cells that are specialized for actuation. 

As in \cite{Nakahira2021}, the diverse components of the immune system are combined in a biological architecture that give a DESS in speed and accuracy. A simple illustration of the stacking of layers and the interplay between engineered and biological systems comes from immune memory and vaccination. Vaccines are slower to develop than adaptive immunity for a new pathogen, but once available, provide relatively fast and accurate protection, a dramatic DESS created by policy, science, medicine, and immune system layers.

IFPs are also present in the immune system, albeit in a more cryptic form because of the ambiguous spatial structure of component interactions. Immune cells move relatively freely through the body, and can have different interactions in different tissue compartments. Nevertheless, if we consider the immune system to be a control system (with, for example, the viral load as the plant), we observe IFPs that are similar in character to those observed in chemotaxis and the nervous system. As in these other systems, we observe a sequential forward loop, and then additional IFPs that modulate the forward loop's responses.

At the cellular level, the forward loop in the antiviral response consists of several components, of which we emphasize antigen presenting cells, CD4+ helper and regulatory T cells, and CD8+ cytotoxic T cells (Fig. \ref{fig:intro_figure_neuro}). Characteristically for the biological systems we consider, each of these categories contains still more sub-categories. 

In the early stages of the antiviral response, antigen presenting cells (APCs) pick up infection-related debris, including proteins manufactured by the pathogen. These APCs then activate T cells with receptors that match the peptides causing the infections; the search for and selection of T cells that are effective against the infection accounts for the delay in this part of the adaptive response. CD8+ T cells can be thought of actuators in the control sense, identifying and killing virus-infected cells so that the virus is unable to replicate and can be removed. However, CD4+ T cells mediate much more complicated dynamics. APCs initially activate CD4+ cells, not CD8+ cells. Indeed, this intermediating activation is \textit{necessary} to allow CD8+ activation. CD4+ cells can suppress or amplify CD8+ cells as well as innate immune and APC circuits, and likewise CD8+ cells can modulate the responses of CD4+ cells and other immune cells \cite{Busse2010,Lund2008,Palm2007}. 

These varied IFPs become more comprehensible when we consider that the immune system is tuned to rapidly respond to changes in tissue health and composition at least as much as, if not more than, it is tuned to the presence of particular pathogen-associated molecular patterns \cite{Pradeu2016}. The textbook forward path model sketched above does not consider this sensitivity to change, but the IFP architectural view potentially does. As the previous case studies have shown, rapid response with limited communication bandwidth can be achieved when signal-dense baselines select and rapidly convey change-related signals for action. Of the three case studies, the immune system is arguably the most opaque; our understanding of it therefore stands to benefit from theory that enables comparisons between systems.

\section{Discussion}
\subsection{Relation to other theories in biology}
The unclear role of IFPs and the related theory gap has attracted attention throughout biology, particularly in neuroscience and in cell signaling. The existence of numerous and diverse IFPs across systems and levels has lead to a sprawl of fragmented explanatory frameworks. These include computation through dynamics \cite{VyasShenoy2020}, recurrent networks \cite{KarDiCarlo2019}, Boolean networks \cite{Kauffman69} Bayesian inference \cite{BastosFriston2012,Swain07}, predictive coding \cite{RaoBallard1999}, and many others. These frameworks have arisen in parallel with the development of methods for increasingly high-throughput and high-resolution measurements of biological systems, which support the idea that internal dynamics are essential for computation and control. However, a crucial level of explanation between circuit motifs of a few components and more phenomenological observations has remained largely unexplored. While each of these theoretical frameworks captures something useful about biological computation and IFPs, they largely neglect component diversity, layers, and most strikingly \textit{speed} or delay, which is crucial in control. The conceptual framework described in this paper identifies these gaps in existing theory, motivating new scalable control approaches and a deeper consideration of IFPs using both old and new control theories \cite{Paper2,Paper3}.

\subsection{Architectures in controllers}

Existing theories from optimal and robust control become computationally intractable even with minimal combinations of SLD+ constraints \cite{Witsenhausen1968,Anderson2019}, except in very special cases  \cite{Paper2}. In contrast, the recently developed parameterization for System Level Synthesis (SLS) explicitly allows for the inclusion of SLD+ constraints on sensing, actuation, \textit{and communication} \cite{Anderson2019, Wang2018TAC, Wang2019TAC}. This allows us to model SATs at various levels of biology. SLS not only produces optimal controllers under SLD+ constraints, but also yields novel controller architectures containing striking similarities (including abundant IFPs) to biological architectures \cite{Paper3}. 

SLS and SLD+ constraints are complementary to the notions of DESS and layered architecture; a full model of biological control includes both, as suggested in Fig. \ref{fig:intro_figure_layers}. Though Fig. \ref{fig:intro_figure_layers} describes important concepts, it abstracts away underlying architectural details, including IFPs. 

IFPs arising from both delay and locality can be loosely interpreted as predictive; in striking recent neuroscience data, they are dominated by the known effects of past and current actions \cite{Musall2019}. The theoretical interpretation is that these IFP signals carry information about how actions propagate through the plant dynamics, which include the body and its environment, as well as about planned future actions, including how communication delays limit such plans and actions. These sources of IFPs encompass more than Bayesian prediction, which is an important conceptual bridge but is a special and ultimately minor source of IFPs. 

In sensorimotor control, there are many more than two layers (trail and bump), two levels (system and nerve), and two tradeoff dimensions (speed and accuracy); these are only first steps in formalizing concepts \cite{Nakahira2021}. Similarly, a theoretical distinction is now reasonably clear between the lack of IFPs here in Fig. \ref{fig:ynfig} and the abundant IFPs in Fig. \ref{fig:intro_figure_neuro} and the companion papers \cite{Paper2, Paper3}. But this too is only a first step in formalizing concepts which will need substantial theoretical and experimental refinement. For example, the vestibulo-ocular reflex, which moves the eye to stabilize gaze despite head motion, could be considered an IFP or an additional control loop with the eye as the plant. Once the sharp distinctions between concepts like layers, levels, and IFPs are clarified with simple examples and theory, a richer theory can develop. At one time, making the distinction between robustness and nominal performance was crucial \cite{Doyle1978}, but ultimately robust performance was an essential overarching concept that did not trivially reduce to its two constituent components \cite{DGZ}. We expect a similar progression here.

\subsection{Towards Experiments and Data} 
By computing optimal SLS controllers with realistic SLD+ constraints, we can design increasingly complicated architectures for which we can precisely explain the function of each component. These optimal controllers may not perfectly match neurobiology, but will lead to more refined experimental directions. Distinct from traditional models, our controller models produce \textit{a priori} models of architecture and behavior from experimentally determined constraints (e.g. bandwidth and delays). Such models allow us to make comparisons with all layers and levels of experimental data in a way that conventional models do not; this will make connections between levels, layers, and behavior more transparent. 

Because SLD+ constraints are ubiquitous in biology, theory gives us a foundation with which to study principles that arise in larger, more flexible systems (like human visuomotor cortex) in smaller, more tractable systems (like \textit{Drosophila}). The evolutionary pressures on organisms are dramatically different, as are many cell- and molecular-level details, but many constraints, and possibly therefore architectures, will be conserved. Our initial survey of IFPs suggests this is indeed the case, but more work in both experimental biology and theory will be needed to establish which architectural features are truly universal.

Central to our goal of uniting biological mechanisms with structured control components is the rapid expansion of detailed measurements of mechanisms that underlie specific behaviors, such as connectomic data in neuroscience and transcriptomic data in biology. These datasets are publicly available, and are a promising avenue to test specific hypotheses about circuits or the cellular composition of the system-level computations. Data about the cell biological level is critically relevant to organismal behavior and physiology, but has only seen limited incorporation into models -- in part because of limitations of traditional theory.
\bibliography{refsp1}
\bibliographystyle{IEEEtran}
\end{document}